\documentclass[aps,prc,onecolumn,preprint,superscriptaddress,showpacs]{revtex4}
\usepackage[dvips]{graphicx}
\usepackage{bm}
\usepackage{amssymb}
\usepackage{amsmath}
\usepackage{nicefrac}
\usepackage{color}

\usepackage{xspace}
\newcommand{\pb}{Papenbrock and Bertsch~\cite{PaBe1998}\xspace}
\newcommand{\dy}{Dyakonov~\cite{Dy1999}\xspace}

\begin{document}
\sloppy

%
% quasiclassical description of bremsstrahlung in $\alpha$ decay
%
\title{Quasiclassical description of bremsstrahlung accompanying\\
$\boldsymbol \alpha$ decay including quadrupole radiation}

\author{U. D. Jentschura}
\affiliation{Max-Planck-Institut f\"ur Kernphysik,
Postfach 103980, 69029 Heidelberg, Germany}
\affiliation{Institut f\"ur Theoretische Physik,
Philosophenweg 16, 69120 Heidelberg, Germany}

\author{A. I. Milstein}
\author{I.~S.~Terekhov}
\affiliation{Max-Planck-Institut f\"ur Kernphysik,
Postfach 103980, 69029 Heidelberg, Germany}
\affiliation{Budker Institute of Nuclear Physics, 630090 Novosibirsk, Russia}

\author{H. Boie}
\author{H. Scheit}
\affiliation{Max-Planck-Institut f\"ur Kernphysik,
Postfach 103980, 69029 Heidelberg, Germany}
\author{D. Schwalm\footnote{present address: 
Weizmann Institute of Science, 76100 Rehovot, Israel}$^{,}$}
\affiliation{Max-Planck-Institut f\"ur Kernphysik,
Postfach 103980, 69029 Heidelberg, Germany}

\begin{abstract}
We present a quasiclassical theory of 
$\alpha$ decay accompanied by bremsstrahlung with a
special emphasis on the case of $^{210}$Po, with the aim of finding
a unified  description that incorporates both the
radiation during the tunneling through the Coulomb wall and the
finite energy $E_\gamma$ of the radiated photon up to $E_\gamma\sim
Q_\alpha/\sqrt{\eta}$, where  $Q_\alpha$ is the $\alpha$-decay
$Q$-value and  $\eta$ is the Sommerfeld parameter. The corrections
with respect to previous quasiclassical investigations are found to
be substantial, and excellent agreement with a full quantum
mechanical treatment is achieved. Furthermore, we find that a
dipole-quadrupole interference significantly changes the
$\alpha$-$\gamma$ angular correlation. We obtain good agreement between 
our theoretical predictions and experimental results.
\end{abstract}


\maketitle

%
% Introduction
%
\section{Introduction}

A characteristic feature of the $\alpha$ decay process is the 
quantum mechanical tunneling~\cite{Ga1929} 
through the so-called Coulomb wall generated by the electrostatic 
interaction of the $\alpha$ particle with the constituent protons of the 
daughter nucleus. Bremsstrahlung in $\alpha$ decay is intriguing because 
of the classically incomprehensible character of radiation emission during 
the tunneling process. Considerable attention has therefore been devoted 
to both 
experimental~\cite{DAEtAl1994brems,KaEtAl1997,Eremin2000,KaEtAl2000,KaEtAl1997jpg} 
as well as theoretical 
investigations~\cite{DyGo1996,PaBe1998,Dy1999,TaEtAl1999,% 
Tk1999,Tk1999jetp,BPZ1999,MaOl2003,MaBe2004}, with the aim of elucidating 
the role of tunneling during the emission process. It is 
necessary to emphasize, however, that the term ``radiation during the 
tunneling process'' has a restricted meaning as the wavelength of the 
photon is much larger than the width of the tunneling region and even 
larger than the main classical acceleration region. It is therefore not 
possible to identify the region where the photon was 
emitted.  Besides, it is possible to write the matrix element of 
bremsstrahlung in different forms using operator identities. As a result, 
the integrands for the matrix element, as well as the relative 
contributions of the regions of integration, will be different depending 
on the operator identities used, although the total answer remains, of 
course, the same. This was demonstrated, e.g., by Tkalya 
in Refs.~\cite{Tk1999,Tk1999jetp}.

In the present paper, we revisit the theory of bremsstrahlung in the
$\alpha$ decay of a nucleus with a special emphasis on the
quasiclassical approximation. The applicability of this
approximation is ensured by the large value of the  Sommerfeld
parameter $\eta$ (see below). We  investigate the range of validity
of the result obtained by \dy and show that it is restricted by the
condition $x \ll 1/\sqrt{\eta}$ where $x=E_\gamma/Q_\alpha$
(here, $E_\gamma$ is the photon energy, and $Q_\alpha$ is the $\alpha$-decay
$Q$-value). Our quasiclassical result has no such a restriction
although we assume $x\ll 1$. It is consistent with the results
of both \dy and \pb
in limiting cases. For the experimentally interesting case
of the $\alpha$ decay of ${}^{210}{\rm Po}$, our result is valid with
high accuracy up to $x \sim 0.1$.

Another subject investigated here is the angular distribution of
emitted photons. The $\alpha$ particle, initially in an $S$ state,
may undergo a dipole transition to a $P$ final state, or a
quadrupole transition to a $D$ state. While the quadrupole
contribution is parametrically suppressed for small photon energies,
the effective charge prefactor for the quadrupole contribution is
large. The dipole-quadrupole interference term vanishes after
angular averaging, but gives a significant contribution to the
differential photon emission probability, resulting in a substantial
deviation from the usually assumed dipole emission characteristics.

Very recently, the results of our high-statistics measurement of 
bremsstrahlung emitted in the $\alpha$ decay of $^{210}$Po have
been published, see Ref.~\cite{ourexp}.
Due to the limited solid-angle coverage of the detectors used in 
this experiment, it was necessary to account for the 
$\alpha$-$\gamma$ angular correlation. Taking into account 
the contributions of the  dipole and 
quadrupole amplitudes in the data analysis, as derived in the present paper, 
good overall agreement between theory and experiment is observed.   

This paper is organized in four sections. In Sec.~\ref{dipole}, we
investigate the leading dipole contribution to the differential
bremsstrahlung probability and  evaluate  the corresponding
amplitude in the quasiclassical approximation. The quadrupole
contribution to the amplitude and its interference with the dipole
part is analyzed in Sec.~\ref{quadrupole}. Conclusions are drawn in
Sec.~\ref{conclu}. Two appendices provide details on the
methods used in the calculations.

%
% DIPOLE EMISSION
%
\section{DIPOLE EMISSION}
\label{dipole}

%
% Emission Probability
%
\subsection{Emission Probability}

It was shown in Ref.~\cite{PaBe1998} that the differential
bremsstrahlung probability $dP/dE_\gamma$ as a function of the energy
$E_\gamma$ of the radiated photon in the dipole approximation has the form
\begin{eqnarray}
\label{basic}
\frac{dP}{dE_{\gamma}} =
\frac{4 \, e^2 Z_{\rm eff}^2}{3\,\mu^2\,E_{\gamma}} \,
\left|{\cal M}\right|^2\,,
\qquad
{\cal M}= \left< R_f \left| \partial_r V \right| R_i\right>\,,
\end{eqnarray}
where natural units with $\hbar=c=\epsilon_0=1$ are applied
throughout the paper,  $e$ is the proton charge and $\mu$ is the
reduced mass of the combined system of $\alpha$ particle and
daughter nucleus,  $V \equiv V(r) = z(Z-z) \alpha/r$ is the
potential of the daughter nucleus felt by the $\alpha$ particle. The
functions $R_i$ and $R_f$ are the radial wave functions of the
initial and final states corresponding to the angular momenta $l=0$
and $l=1$, respectively (see App.~\ref{radial}). The effective
charge for a dipole interaction between an $\alpha$
particle with charge number $z=2$ and mass number $4$ emitted from a
parent nucleus with charge number $Z$ and mass number $A$ is (see
also App.~\ref{effcharge})
\begin{equation}
\label{Z1eff}
Z_{\rm eff}=Z^{(1)}_{\rm eff} \approx \frac{2\,A - 4\,Z}{A} = \frac25\,,
\end{equation}
where the latter value is relevant for the experimentally
interesting case of the $\alpha$ decay of ${}^{210}{\rm Po}$ ($Z = 84$).
Evaluating the effective charge with accurate 
values for the masses of the alpha particle and
the daughter nucleus ($^{206}$Pb, $Z = 82$), as given in
Ref.~\cite{Fi1996} yields a value of $Z^{(1)}_{\rm eff}= 0.399$.

In the present paper, we calculate the matrix element ${\cal M}$ in
the quasiclassical approximation taking into account the first
correction of the order ${\cal O}(\eta^{-1})$, and the corresponding
result is given below in Eq.~(\ref{Our}). However, before presenting
and discussing our formula for the matrix element ${\cal M}$,
let us briefly review several results for ${\cal M}$ obtained earlier in the
quasiclassical approximation. These are illustrative with respect to
their range of applicability and with respect to the importance of the
tunneling contribution.

%
% Dipole Transition Matrix Element
%
\subsection{Dipole Transition Matrix Element}

Various approximations have been applied for the evaluation of the matrix
element $\cal M$ in Eq.~(\ref{basic})
\cite{Dy1999,PaBe1998,TaEtAl1999,Tk1999,Tk1999jetp}.
The approximations are intertwined with the identification of particular
contributions to the real and imaginary parts of the matrix element
$\mathcal M$ due to ``tunneling'' and due to ``classical motion'' of the
$\alpha$ particle.

We  use the convention that the  complex phase of the matrix element
$\cal M$ should be  chosen in such a way that it becomes purely real
in the classical limit $E_\gamma \to 0$. Our definition of $\cal M$
is consistent with that used in Ref.~\cite{Dy1999} and differs by a
factor $\rm{i}$ from the definition used in Ref.~\cite{PaBe1998}.

Equation~(5) in the work of \pb contains a fully quantum mechanical
result for the photon emission amplitude $\mathcal{M}$, expressed in
terms of regular and irregular Coulomb functions, without any
quasiclassical approximations. However, the physical interpretation
of this result depends on a comparison with a
quasiclassical approximation, as only such a comparison clearly
displays the importance of the finite photon energy and the emission
amplitude during tunneling. \pb therefore present and discuss 
a quasiclassical
expression for the imaginary part of their matrix element (real part
for our convention), ignoring the contribution from the tunneling
process to the emission amplitude. Note that the quasiclassical
expression of \pb provides a very good approximation
for the imaginary part of their matrix element up to very large photon
energies with $x\sim 0.6$.

In contrast, the quasiclassical result of \dy is valid only for very
small photon energies ($x\ll 1/\sqrt{\eta}$), but includes
contributions from tunneling.
Here, we unify the treatments of Refs.~\cite{PaBe1998,Dy1999} and obtain a
quasiclassical differential emission  probability $dP/dE_\gamma$,
which includes the effect of photon emission during the tunneling
process and which is substantially more accurate for higher photon
energies than that of \dy.

The  quasiclassical  approximation for the  wave functions of the
system of an $\alpha$ particle and a daughter nucleus in the initial
and final states is valid for large values of the Sommerfeld
parameters $\eta_{i,f}=z(Z-z)e^2\mu/k_{i,f}$ with $k_i=\sqrt{2\mu
Q_\alpha}$ and $k_f=\sqrt{2\mu(Q_\alpha-E_\gamma)}$. The indices $i$
and $f$ are reserved for initial and final configurations throughout
the paper. The value of $\eta_i$ for the decay of $^{210}$Po, which
is the experimentally most interesting nucleus, is $22.0$,
while the $Q$-value is $Q_\alpha = 5.40746 \, {\rm MeV}$~\cite{Fi1996}.

If one neglects the contribution to the matrix element $\mathcal M$
from the region $r\lesssim r_0$, where $r_0$ is the nuclear radius,
then one is consistent with the simple assumption for the potential as
a square well for $r < r_0$ and a pure repulsive Coulomb potential for
$r > r_0$.  Implementing this procedure according to \pb, one obtains
the following approximation for the 
real part $M$ of the matrix element of $\mathcal M$,
\begin{align}
\label{Pap_matix_el_exact}
M = \mbox{Re}\, \cal M \approx &
\,  \eta_i\,\sqrt{\frac{2\,k_i}{\pi\, k_f}}\, \nonumber\\
& \; \times \int_{0}^{\infty}\frac{dr}{r^2} \,
  F_1(\eta_f,k_f r)\, F_0(\eta_i,k_i r) \,.
\end{align}
Here, $F_0$ and $F_1$ are the regular Coulomb radial wave functions
corresponding to angular momenta $l=0$ and $l=1$, respectively.
The importance of the contribution of the region $r \lesssim r_0$ was
discussed in Refs.~\cite{PaBe1998,Dy1999}. For relatively small photon
energies, which are interesting from an experimental point of view,
this contribution is not significant, and we do not consider it in the
present paper.  For high photon energies, the contribution of the
region $r \sim r_0$ can be important (see Ref.~\cite{BPZ1999}).
The explicit form of $M$ is given by Eqs.~(6) and (7) of
Ref.~\cite{PaBe1998,PaBe1998-misprint}.

We have calculated the integral 
in Eq.~(\ref{Pap_matix_el_exact})
using quasiclassical wave functions, keeping  the correction of order
${\cal O}(\bar{\eta}^{-1})$ and ignoring terms of order ${\cal
O}(1/\bar{\eta}^2)$ and higher in the expansion for large ``mean''
Sommerfeld parameter ${\bar \eta} = (\eta_i + \eta_f)/2$. The result
of such a quasiclassical calculation for the real part of $\cal M$ 
reads
\begin{align}
  \label{Pap_matix_el_q}
  \widetilde{M} =&
  \sqrt{\frac{2\,k_i}{\pi\, k_f}}\,
  \frac{k_i\,k_f}{k_i+k_f}\,
  \frac{\eta_i}{\bar{\eta}} \nonumber\\
  & \times
  \xi \, {\rm e}^{-\pi \xi /2} \,
  \left[-K_{{\rm i}\xi}'(\xi)-
    \frac{1}{\bar{\eta}} \,
    K_{{\rm i}\xi}(\xi)\right]\,,
\end{align}
where $\xi=\eta_f-\eta_i$, and
$K_{a}(b)$ is the modified Bessel function. The derivative is
$K_{a}'(b)= \frac{\partial}{\partial b}K_{a}(b)$. \pb also
calculated the real part of the integral (\ref{Pap_matix_el_exact})
in the quasiclassical approximation. Note, however, that 
their term of order ${\cal O}(\bar{\eta}^{-1})$ [see Eq.\ (14) of 
Ref.~\cite{PaBe1998}]
contains an additional factor $\sqrt{3}/2$ in comparison to 
Eq.~(\ref{Pap_matix_el_q}). 

\begin{figure}
\centering
\includegraphics[scale=0.7]{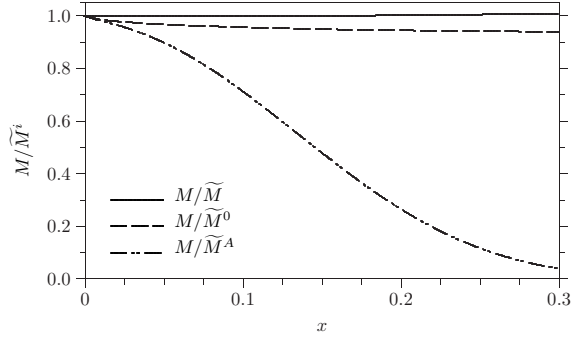}
\caption{\label{largex} The ratio of $M = \mbox{Re}\, \cal M$, 
see Eq.~(\ref{Pap_matix_el_exact}), to various
approximations, as a function of $x=E_\gamma/Q_\alpha$ for 
for the case of $^{210}$Po ($\eta_i =22.0$). The solid line shows 
$M/\widetilde{M}$ with $\widetilde{M}$ taken from
Eq.~(\ref{Pap_matix_el_q}), illustrating the excellent agreement of
the quasiclassical matrix element with the exact result in the range
$x \leq 0.3$. For the dashed line we use 
$\widetilde{M}^0$,
which is obtained from $\widetilde{M}$ by omitting the
correction of order ${\cal O}(\bar{\eta}^{-1})$, showing a deviation
of less than $5\%$. For the dash-dotted line the asymptotics
$\widetilde{M}^A$ as given in Eq.~(\ref{MPBA}) is used. The deviation of the
latter curve from the others illustrates that the 
``$x \to 0$''-asymptotics is indeed only valid for $x \ll 1/\sqrt{\eta}$.}
\end{figure}

Figure~\ref{largex} shows the ratio ${\rm Re} \, {\cal M}/ \widetilde{M} = 
M/\widetilde{M}$ as a function of $x$ for $\eta_i=22.0$,
corresponding to the case of $^{210}$Po. Note that even at $x=0.3$,
the deviation of the quasiclassical result with the correction
${\cal O}(\bar{\eta}^{-1})$ taken into account is less than $1\%$
(solid line), while without this correction 
($\widetilde{M}^0$), the deviation is about $5\%$ (dashed line).

Strictly speaking, the validity of the evaluation of the transition
matrix element (\ref{basic}) with the wave functions taken in the
quasiclassical approximation requires special consideration (see \S
51 of Ref.~\cite{LaLi1958}) because of possibly noticeable
contributions from the vicinities of the turning points. However,
one can show that, for $\xi\ll\bar\eta$ (or $x \ll 1$), the
contributions of the vicinities of the classical turning points
($r_{ci} =2\eta_i/k_i$ and $r_{cf} =2\eta_f/k_f$ in our case) are
small. For $x\lesssim 1$, these contributions are no longer negligible.

For  $x\ll 1$,  we have $\xi=\eta_i\,(x/2+3x^2/8+\ldots)\,$. If
$x^2\,\eta_i\ll 1$ (even if $x \, \eta_i \lesssim 1$), we can
replace in (\ref{Pap_matix_el_q}) $\xi$ by $x\eta_i/2$ and make the
substitution $\eta_f \to \eta_i$ and $k_f \to k_i$. As a result, we
obtain the following asymptotics of  $\widetilde{M}$:
\begin{align}
\label{MPBA} 
\widetilde{M}^A &=\frac{k_i}{\sqrt{2 \pi}} \,
\left(\frac{x \eta_i}{2}\right) \, {\rm e}^{-\pi x \eta_i/4} \,
\nonumber\\
& \; \times \left[-K_{{\rm i} x \eta_i/2}'\left(\frac{x
\eta_i}{2}\right) -\frac{1}{\eta_i} \, K_{{\rm i} x
\eta_i/2}\left(\frac{x \eta_i}{2}\right)\right]\,.
\end{align}
From the dash-dotted line of Fig.~\ref{largex}, we see that the
ratio $M/\widetilde{M}^A$ deviates
substantially from unity, illustrating that the applicability of
Eq.~(\ref{MPBA}) is indeed restricted to very small values of $x$.

Since
\begin{equation}
K_{i\nu}(x) = \exp(-\pi\nu/2) \int_0^\infty \cos(x\sinh t - \nu t)\,dt\,,
\end{equation}
the quantity $\widetilde{M}^A$ without the ${\cal
O}(\eta_i^{-1})$ correction exactly coincides with the real part of
the amplitude ${\widetilde{\cal M}}_{\rm D}$ obtained by \dy, 
where
\begin{subequations}
\label{MD}
\begin{align}
\label{MDM}
\widetilde{\cal M}_{\rm D} =& \;
\frac{k_i}{\sqrt{2\pi}} \, J_{\rm D}\left(\frac{x\,\eta_i}{2}\right)\,,\\
\label{MDJ}
J_{\rm D}(y) =& \; -{\rm i}\, y \exp(-\pi y)\nonumber\\
&\times\int_{0}^{\infty} dt \,\sinh(t) \, \exp[{\rm i} \, y \,
(\sinh t-t)]\,,
\end{align}
\end{subequations}
showing that this result is applicable only to very small photon
energies. $\widetilde{\cal M}_{\rm D}$
contains both a real and an imaginary part and thus takes
photon emission during tunneling into account. It was pointed out by
\dy that the imaginary part of $\widetilde{\cal M}_{\rm D}$ at $x\eta_i \sim
1$ is of the same order as the real part. For instance, in the case
of the $\alpha$ decay of $^{210}$Po with $\eta_i=22.0$, we have
$ x \, \eta_i \sim 1$ for $x\approx 0.05$. This indicates
that the imaginary part of $\widetilde{\cal M}_{\rm D}$ is important.

Our quasiclassical approximation $\widetilde{\cal M}$ for the dipole 
transition matrix element $\cal M$ from Eq. (\ref{basic}),
which includes the contributions from the tunneling of the $\alpha$
particle (we would like to refer to this result as the ``unified
result'' in the following sections of the current paper) has the
form:
\begin{subequations}
\label{Our}
\begin{align}
\label{OurA} 
\widetilde{\cal M} =& \sqrt{\frac{2\,k_i}{\pi\, k_f}}\,
\frac{k_i\,k_f}{k_i+k_f}\, \frac{\eta_i}{\bar{\eta}}\,\left[J(\xi) +
  \frac 1{\bar{\eta}} J_1(\xi)\right]\,,\\
\label{OurJ}
J(y) =& \; {\rm i}\, y \exp(-\pi y)\nonumber\\
&\times\int_{0}^{\infty} dt \,\sinh(t) \, \exp[{\rm i} \, y \,
(t-\sinh t)]\,,\\
\label{OurB} J_1(y)= &- y \exp(-\pi y) \,\int_{0}^{\infty} dt
\,\exp[{\rm i} \, y \, (t - \sinh t)]\,.
\end{align}
\end{subequations}
The derivation of Eqs.~\ref{Our} (see App.~\ref{radial} for
more details) involves a shift of the integration region into the
complex plane, which is needed to take the tunneling region
into account in a quasi-classical treatment, as explained 
in~\cite{DyGo1996,Dy1999}.
Note that $J(y)$ is the complex
conjugation of the function $J_{\rm D}(y)$ as defined in
Eq.~(\ref{MD}). Although this is irrelevant for the calculation of the
bremsstrahlung emission probability, it is important for the
dipole-quadrupole interference term discussed in
Sec.~\ref{quadrupole}.  The region of applicability of Eq.~(\ref{Our})
is much wider than that of Eq.~(\ref{MD}), because at small $x$ there
is no additional restriction $x\ll 1/\sqrt{\eta}$ which may otherwise
constitute a strong limitation at large value of $\eta$, as it was
shown above. Moreover, Eq.~(\ref{Our}) contains a correction ${\cal
  O}(\bar\eta^{-1})$, which is also essential.

\begin{figure}
\centering
\includegraphics[scale=0.7]{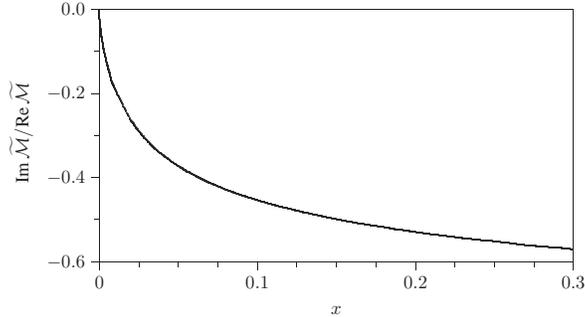}
\caption{The ratio $\mbox{Im}\,\widetilde{\cal M}/
\mbox{Re}\,\widetilde{\cal M}$ for the case of 
$^{210}$Po ($\eta_i = 22.0$). The graph illustrates that
the imaginary part of the matrix element $\widetilde{\cal M}$
is quite substantial even for moderate values of $x$.}
\label{ImRe}
\end{figure}

\begin{figure}
\centering
\includegraphics[width=8cm]{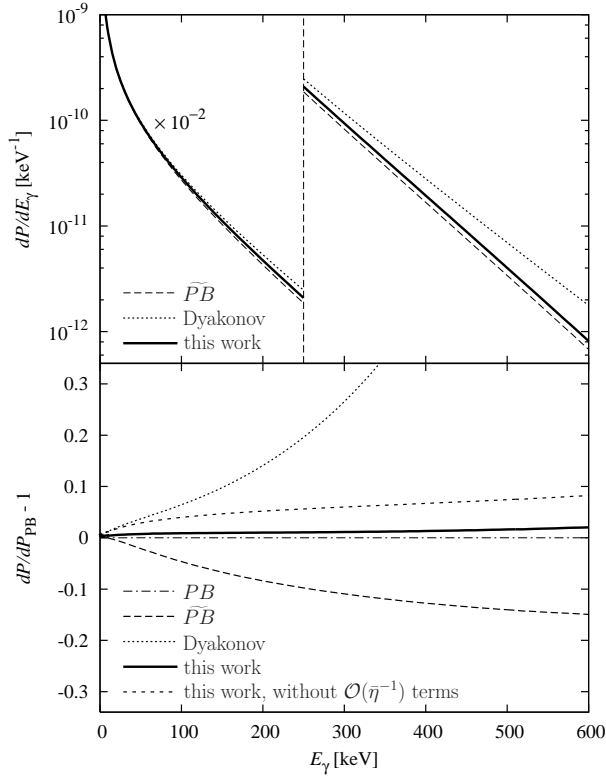}
\caption{\label{fig3}
Differential branching ratio $dP/dE_\gamma$ for bremsstrahlung emission during
$\alpha$ decay of $^{210}$Po in units of inverse keV (top panel)
and relative to the fully quantum mechanical calculation (PB)
of \pb (bottom panel).
The thick solid curve corresponds to our quasiclassical result,
as given in Eq.~(\ref{Our}). The other results are based on the 
quasiclassical treatment of Ref.~\cite{PaBe1998} 
($\widetilde{\rm PB}$), the semiclassical treatment of 
Dyakonov~\cite{Dy1999}, and on Eq.~(\ref{Our}) 
of the present work but neglecting the correction terms of 
order ${\cal O}(\bar \eta^{-1})$.}
\end{figure}

Our unified result (\ref{Our}) can be
used with high accuracy up to $x\sim 0.3$.  
The real part of $\widetilde{\cal M}$ is identical 
to $\widetilde{M}$ given in Eq.~(\ref{Pap_matix_el_q}) and is 
discussed already in detail above (see also Fig.~\ref{largex}).
Figure~\ref{ImRe} shows the ratio 
$\mbox{Im}\,\widetilde{\cal M}/\mbox{Re}\,\widetilde{\cal M} = 
\mbox{Im}\,\widetilde{\cal M}/ \widetilde{M}$ 
as a function of $x$ at $\eta_i=22.0$,
i.e.~for the case of $^{210}{\rm Po}$. One
can see that $\mbox{Im}\,\widetilde{\cal M}$ is not small in
comparison to $\mbox{Re}\,\widetilde{\cal M}$. Thus, the imaginary 
part gives a noticeable
contribution to $dP/dE_\gamma$ even for small $x$, and should not be
neglected. This point was also emphasized in
Refs.~\cite{Dy1999,TaEtAl1999}.

%
% Quantitative comparison of various quasiclassical results
%
\subsection{Quantitative comparison of various quasiclassical results}

As a last step, we compare in Fig.\ \ref{fig3} the differential
bremsstrahlung probability $dP/dE_\gamma$ 
for the bremsstrahlung accompanying $\alpha$ decay of $^{210}$Po
obtained with the use of
our matrix element $\widetilde{\cal M}$ given in Eq.~(\ref{Our}), 
the matrix element $\widetilde{M}$ [Eq.~(\ref{Pap_matix_el_q})], 
corresponding to the quasiclassical approach of 
Ref.~\cite{PaBe1998}, and
the matrix element $\widetilde{\cal M}_{\rm D}$ [Eq.~(\ref{MD})], to the
full quantum mechanical formula given by Eq.\ (5) in
\cite{PaBe1998}. A detailed comparison is shown in the bottom panel
of Fig.\ \ref{fig3} in the $\gamma$ energy range
$0<E_\gamma<600$~keV. While the result of \dy (dotted line) deviates
roughly by a factor two at $E_\gamma=600$~keV, our result
(thick solid line) agrees with the exact quantum mechanical
treatment within about 2~\% at this photon energy; 
the inclusion of the 
${\cal O}(\bar \eta^{-1})$ correction is crucial in obtaining
this agreement, as is evident from the supplementary curve in the
bottom panel, where we omit the $J_1$ term from Eq.~(\ref{OurB}).
The quasiclassical approximation of \pb, obtained by neglecting  the
imaginary part of the matrix element (dashed line), deviates by more
than 15~\% at $E_\gamma=600$~keV. 
As expected, at low photon energies
all results agree with each other.

\begin{figure}
\centering
\includegraphics[width=8cm]{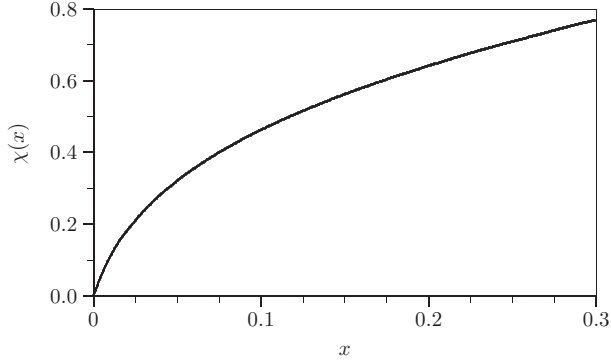}
\caption{\label{fig4} Interference term $\chi(x)$,
defined in Eq.~(\ref{chi}), for the bremsstrahlung accompanying
$\alpha$ decay of $^{210}$Po.}
\end{figure}

%
% QUADRUPOLE EMISSION
%
\section{QUADRUPOLE EMISSION}
\label{quadrupole}

We are now concerned with the quadrupole component of the
bremsstrahlung  probability and the angular distribution of the
radiation due to interference with the dipole components. The
outgoing $\alpha$ particle defines an axis of symmetry.
We therefore we may use $d\Omega = 2\pi \sin\theta d\theta$
in order to describe the solid angle element of the 
photon spanning an infinitesimal range of polar angles 
$\theta$ with respect to the direction of the emitted 
$\alpha$-particle. We assume that a
summation with respect to photon polarization is performed. Within
the dipole approximation, Eq.~(\ref{basic}) gives rise to an angular
distribution of the form
\begin{eqnarray}
\left. \frac{d^2P}{dE_{\gamma} d\Omega} \right|_{\rm dip} =
\frac{e^2 \left( Z^{(1)}_{\rm eff} \right)^2
\sin^2\theta}{\pi\,\mu^2\,E_{\gamma}} \,
\left|{\cal M}\right|^2\,,
\end{eqnarray}
where the index refers to the dipole approximation.
Including the quadrupole term (see  App.~\ref{radial}), this
formula should be generalized to
\begin{align}
\label{genangle}
\frac{d^2P}{dE_{\gamma} d\Omega} =&\;
\frac{e^2 \sin^2\theta}{\pi\mu^2E_{\gamma}}
\left|Z_{\rm eff}^{(1)} \, {\rm e}^{{\rm i}\delta_1} \,
{\cal M} \, +
Z_{\rm eff}^{(2)} \,
{\rm e}^{{\rm i}\delta_2} \,
{\cal N} \, \cos\theta
\right|^2
\nonumber\\[2ex]
\,= & \; \left. \frac{d^2P}{dE_{\gamma} d\Omega} \right|_{\rm dip} \,
[1+\chi\,\cos\theta\,] +{\cal O}({\cal N}^2)\, ,\\
\intertext{with}
\label{chi} \chi=& \; 2 \, \frac{Z_{\rm eff}^{(2)}}{Z_{\rm eff}^{(1)}}\,
{\rm Re} \left( \frac{{\cal M} \, {\cal N}^*}{|{\cal M}|^2} \, {\rm
e}^{{\rm i} (\delta_1 - \delta_2)} \right) \,,
\end{align}
where $\delta_1$ and $\delta_2$ are the Coulomb phases corresponding
to the angular momenta  $l=1$ and $l=2$, respectively. The effective
quadrupole charge $Z^{(2)}_{\rm eff}$ is approximately given by (see
App.~\ref{effcharge})
\begin{equation}
\label{Z2eff} Z_{\rm eff}^{(2)} 
\approx 2 + \frac{16 (Z-A)}{A^2} = 1.954\,
\end{equation}
for the  case of $^{210}{\rm Po}$, and is roughly five times larger
than the dipole effective charge given in Eq.~(\ref{Z1eff}). 
Exact masses~\cite{Fi1996} lead to the same
result $Z_{\rm eff}^{(2)}= 1.954$ (up to the last decimal digit indicated).

Calculating the quadrupole matrix element ${\cal
N}$ within the quasiclassical approximation and keeping the leading
in $1/\bar{\eta}$ term, we obtain
\begin{subequations}\label{OurQ}
\begin{align}
\widetilde{\cal N} =& \;- \sqrt{\frac{2\,k_i}{\pi\, k_f}}\,
\frac{k_i\,k_f}{k_i+k_f}\,
\frac{\eta_i}{\bar{\eta}}\,v\,J_1(\xi) \,.
\end{align}
\end{subequations}
We here neglect a parametrically suppressed
next-to-leading order correction to the 
interference term, discussed in more detail in Appendix~\ref{radial},
and we introduce the notation $v = \sqrt{2 Q_\alpha/\mu}$, where $v$
approximately equals
the final velocity of the $\alpha$ particle for bremsstrahlung emission
with $x \ll 1$. For $^{210}{\rm Po}$, we have $v
\approx 0.05$. Note that for a large Sommerfeld parameter $\eta_f$ we
have
\begin{equation}\label{phases}
{\rm e}^{{\rm i}\,(\delta_2 - \delta_1)} \approx {\rm i} +
\frac{2}{\eta_f} \, .
\end{equation}
Because of this ``${\rm i}$'' in the right-hand side~of
Eq.~(\ref{phases}), the imaginary parts of the functions $J(\xi)$
and $J_1(\xi)$ become very important for the interference term. The
quantity $\chi(x)$ defined in Eq.~(\ref{chi}), which determines the relative
magnitude of the interference term, vanishes for $x \to 0$ in the leading
in $1/\bar\eta$ approximation as
\begin{equation}
\chi\approx  \frac{\pi}{2} \, 
\frac{Z_{\rm eff}^{(2)}}{Z_{\rm eff}^{(1)}} \, v \, \eta_i\,x\,.
\end{equation}
For ${}^{210}{\rm Po}$, this evaluates to $\chi\approx 8.64\, x$.
Therefore, the value of the coefficient $\chi(x)$ becomes
significant already at very small photon energies ($\chi\sim 0.1$ at
$E_\gamma\sim 0.06$~MeV). The coefficient $\chi(x)$,
calculated with our quasiclassical results 
$\widetilde{\cal M}$ and $\widetilde{\cal N}$, is plotted in
Fig.~\ref{fig4} as a function of $x = E_\gamma/Q_\alpha$ for the
experimentally interesting case of ${}^{210}{\rm Po}$.

When integrating Eq.~(\ref{genangle}) over the total solid 
angle, the interference term drops out and the remaining 
quadrupole contribution to the differential emission 
probability is of the order ${\cal O}( \widetilde {\cal N}^2)$; 
for the case of $^{210}$Po this contribution amounts to less than 
1.5\,\% of the leading dipole term for photon energies up to 
600 keV ($x \approx 0.1$).

%
% Summary
%
\section{Summary}
\label{conclu}

In summary, using a quasiclassical approximation we have obtained an
expression  for the dipole bremsstrahlung probability during $\alpha$
decay which is in agreement with the full quantum mechanical treatment
of \pb in a substantially wider region
of the variable $x=E_\gamma/Q_\alpha$ in comparison with all
previous quasiclassical results. Our amplitude given in
Eq.~(\ref{Our}) contains both real and imaginary parts and,
thus, includes the contribution from bremsstrahlung during the
tunneling process. Our results demonstrate that the latter contribution is
not negligible even for rather small $x$. As an illustration of
these statements, we have considered the experimentally important
case of $^{210}$Po. 

Furthermore, we find the quasiclassical expression for the
contribution of the interference term of the dipole and quadrupole
components to the double differential bremsstrahlung probability
(with respect to the energy and the solid angle of the photon). This
contribution turns out to be significant.
Because of obvious limitations to the solid angle that can be
covered by detectors in a realistic experiment, 
the angular distribution needs to be considered in the analysis of the
experimental data, even though the quadrupole
term makes a negligible contribution to the bremsstrahlung probability
after integration over the entire solid angle. Using the expression for the 
dipole and quadrupole amplitudes presented in this work, the data analysis 
in our recent experiment was performed as described in Ref.~\cite{ourexp},
and good overall agreement of our theoretical and experimental results 
was obtained (see Fig.~5 of Ref.~\cite{ourexp}). We note, however, 
that a certain subtle question remains with respect to a 
next-to-leading order, nuclear model-dependent 
correction to the dipole-quadrupole interference
term, as discussed in Appendix~\ref{radial}. These questions leave room for 
further interesting investigations in the context  
of under-the-barrier emission of bremsstrahlung in $\alpha$ decay
in the future.

\begin{acknowledgments}

A.I.M. and I.S.T. gratefully acknowledge the School of Physics at
the University of New South Wales, and the Max-Planck-Institute for
Nuclear Physics, Heidelberg, for warm hospitality and support during
a visit. D.S.~acknowledges support by the 
Weizmann Institute through the Joseph Meyerhoff program,
and U.D.J.~acknowledges support from the Deutsche
Forschungsgemeinschaft (Heisenberg program). The work was also
supported by RFBR Grant No. 05-02-16079. \end{acknowledgments}

\appendix

%
% Effective charges
%
\section{Effective charges}
\label{effcharge}

The purpose of this appendix (see also \cite{Eisenberg}) is to
clarify how the effective charges in Eqs.~(\ref{Z1eff}) and
(\ref{Z2eff}) for the dipole and the quadrupole terms arise in the
interaction of a two-body system (charges $eZ_1$ and $eZ_2$,  masses
$m_1$ and $m_2$, $e$ is the proton charge) with a photon of
polarization $\bm{\epsilon}$ and wave vector $\bm{q}$, as given by
the part of the Hamiltonian corresponding to emission of a photon,
\begin{align}
\label{HI}  H_I =& - \frac{eZ_1}{m_1}\,\bm{\epsilon}^*\cdot\bm{p}_1
\,
\exp\left( -{\rm i} \, \bm{q}\cdot\bm{r}_1\right)\nonumber\\
& - \frac{eZ_2}{m_2}\, \bm{\epsilon}^*\cdot {\bm{p}}_2 \, \exp\left(
-{\rm i} \, \bm{q}\cdot\bm{r}_2\right)\, .
\end{align}
We define the total mass $M = m_1+m_2$, the reduced mass $\mu = m_1
m_2/M$ , the center-of-mass coordinate $\bm{R} = (m_1/M) \bm{r}_1 +
(m_2/M) \bm{r}_2$, and the relative coordinate $\bm{r} = \bm{r}_1 -
\bm{r}_2$.  Let $\bm p$ and $\bm P$ be the momenta corresponding to
the coordinates $\bm r$ and $\bm R$, respectively. Then $\bm{p}_1
=\bm p+ (m_1/M)\bm P$ and $\bm{p}_2 = -\bm p+ (m_2/M)\bm P$. Writing
the Hamiltonian (\ref{HI}) in the center-of-mass frame ($\bm P=0$)
and performing its expansion  over $|\bm q\cdot\bm r|\ll 1$ up to
the first term, we obtain
\begin{equation}
\label{HIRES} H_I = -e \frac{\bm{\epsilon}^*\cdot\bm{p}}{\mu}\,
\left[ Z^{(1)}_{\rm eff} -{\rm i} \, Z^{(2)}_{\rm eff} \,
\bm{q}\cdot\bm{r} \right] {\rm e}^{-{\rm i}\bm{q}\cdot\bm{R}}\,,
\end{equation}
where the overall phase factor ${\rm e}^{-{\rm i}\bm{q}\cdot\bm{R}}$
can be safely ignored.

The effective charges are
\begin{align}
Z^{(1)}_{\rm eff}
=& \frac{Z_1 \, m_2 - Z_2 \, m_1}{m_1 + m_2} \,,
\\[2ex]
Z^{(2)}_{\rm eff} =&
\frac{Z_1 \, m_2^2 + Z_2 \, m_1^2}{(m_1 + m_2)^2} \,.
\end{align}

%
% Matrix elements
%
\section{Matrix elements}
\label{radial}

In this appendix we present some details of the
derivation of our quasiclassical dipole (\ref{Our}) and quadrupole
(\ref{OurQ}) matrix elements. The wave function of the final state
has the form (see, e.~g., \S 136 and \S 137 of~\cite{LaLi1958}):
\begin{align}
\psi_f(\bm{r})&\equiv \psi^{(-)}_{\bm{k}_f}(\bm{r})\nonumber\\
=&
\frac{1}{2k_f} 
\sum_{l=0}^\infty {\rm i}^l \, {\rm e}^{-{\rm i}\delta_l}\, 
(2 l + 1)\, R_{k_f,\,l}(r) \, P_l(\bm{n} \cdot
\bm{\lambda})\,,
\end{align}
where $\bm n=\bm r/r$, $\bm \lambda=\bm k_f/k_f$, $P_l$ are the
Legendre polynomials, and
\begin{equation}
\label{finalstate}
R_{k_f,\,l}(r)= \frac{2}{r}F_l(\eta_f, k_f r)
\end{equation}
is the regular radial solution of the
Schr\"{o}dinger equation in a Coulomb field. 

When comparing to Eq.~(5) of Ref.~\cite{PaBe1998}, 
it is evident that the ansatz (\ref{finalstate}) 
for the final-state wave function
corresponds to the neglect of the contribution from the 
irregular solution in the Couloumb field, which 
given by the term $2 r^{-1} G_l(\eta_f, k_f r) \, \tan\alpha$
in the integral in Eq.~(5) of Ref.~\cite{PaBe1998}.
The basis for our approximation  (\ref{finalstate})
is as follows.
The asymptotics of the wave function at $kr \gg 1$ is
$2 r^{-1} \sin(k r + f_C + \delta_N)$, where $f_C$
corresponds to the asymptotics in a pure Coulomb field, and 
the phase shift
$\delta_N$ is due to the nuclear potential,
which has a size of the order of $r_0$.
So, the asymptotic form of $R_{k_f, l}(r)$  is
proportional to  
$\cos(\delta_N)F_l(\eta_f, k_f r) +
\sin(\delta_N)G_l(\eta_f, k_f r)$. 

It follows from the
quasiclassical approximation that the wave function at 
$r \ll r_{cf} \equiv r_t$,
where $r_t=2 \eta_f/(\mu v)$ is the classical turning point,
is given by $F_l(\eta_f, k_f r) \propto 
\exp[-2\sqrt{2\mu v\eta_f} \, (\sqrt{r_t}-\sqrt{r})]$. Therefore,
$\delta_N\propto \exp[-8 \eta_f] \ll 1$, and one is therefore
led to the tentative conclusion that the term with the $G$ function
should be entirely negligible.
However, one might still object that the function $G$ is
exponentially large at $r \ll r_t$, namely
$G_l(\eta_f, k_f r)\propto\exp[2\sqrt{2\mu v\eta_f}\,(\sqrt{r_t}-\sqrt{r})]$.
In order to convince ourselves that the contribution
from the $G$ functions is indeed small,
let us consider as an example the term $\sin(\delta_N)G_0(\eta_i, k_i r) \, 
G_l(\eta_f, k_f r)$. This term is proportional to
$\exp[-4(\eta_f-\eta_i)]\exp(-4\sqrt{2 \mu v \bar\eta}\sqrt{r})$. Therefore,
the main contribution to the transition 
matrix element from this term is given
by the region $r\sim 1/(32 \mu \eta v)=1/(32 \mu(zZ\alpha)) \ll r_0$.
This contribution is indeed small, and 
we can safely use the approximation (\ref{finalstate})
for the final state of the $\alpha$ particle.

The wave function of
the initial state is given by an $S$ state which consists of
an outgoing wave at $r\to\infty$,
\begin{equation}
\psi_i(\bm r) = \frac{R_0(r)}{\sqrt{4\pi}}=\frac{1}{\sqrt{4\pi}\,
r}\,[G_0(\eta_i,k_i r)+{\rm i}\,F_0(\eta_i,k_i r)]\, .
\end{equation}
Substituting the wave functions into the  transition matrix element
and taking the integrals over the angular parts of the 
$\alpha$ particle wave functions, we obtain
\begin{align}
&\left< f | H_I | i \right> = -  \frac{\sqrt{4 \pi}\,e\,
\eta_i\,k_i}{2k_f\,\mu^2
\omega}\, \bm{\epsilon}\cdot\bm{\lambda} \nonumber\\
&\times\Bigl\{ Z^{(1)}_{\rm eff} e^{{\rm i}\delta_1}
\int\limits_0^\infty {\rm d}r R_{k_f,\,1}(r)
R_{0}(r)\nonumber\\
& + Z^{(2)}_{\rm eff} e^{{\rm i}\delta_2}\,
\cos\theta\,\int\limits_0^\infty  {\rm d}r R_{k_f,\,2}(r) 
\left[ r \omega + \frac{9}{\mu r} \right] R_{0}(r) \Bigr\}\, ,
 \label{vogts}
\end{align}
where $\theta$ is the angle between the vectors $\bm q$ and $\bm
\lambda$, $\omega = |\bm{q}| = E_\gamma$. Defining ${\cal M}$ and
${\cal N}$ as
\begin{align}
{\cal M} =& -{\rm i}\eta_i \sqrt{\frac{k_i}{2\pi k_f}}\int_0^\infty
dr R_{k_f,\,1}(r)R_{0}(r)\,,
\\
{\cal N} =& -{\rm i}\eta_i \sqrt{\frac{k_i}{2\pi k_f}}\,
\int_0^\infty dr  R_{k_f,\,2}(r)\left(\frac{9}{\mu r} + r \omega
\right) R_{0}(r)\,,\label{MN}
\end{align}
and taking into account that the sum over the photon polarizations
gives $ \sum |\bm \epsilon \cdot \bm \lambda|^2=\sin^2\theta $, we
obtain the double-differential bremsstrahlung probability as
\begin{align}
\frac{d^2P}{dE_{\gamma} d\Omega} =&
\frac{e^2}{\pi\mu^2E_{\gamma}}\sin^2\theta\, |{\cal C}|^2 \,
,\nonumber\\
{\cal C} =&  Z^{(1)}_{\rm eff} e^{{\rm i}\delta_1} {\cal M} +
Z^{(2)}_{\rm eff} e^{{\rm i}\delta_2} {\cal N} \cos\theta \,,
\end{align}
in agreement with Eq.~(\ref{genangle}). Then we use  the
quasiclassical approximation for the radial part of the wave
functions (see, e.g., \S~48, 49 of Ref.~\cite{LaLi1958}). The matrix
elements with the quasiclassical  radial wave functions have been
calculated using methods described in detail in
Ref.~\cite{Alder56}. Although these methods are in principle well known,
we present here some details of the calculation.

Let us consider first the contribution $I_0$ of the classically 
allowed region to the matrix element
\begin{equation}
I_1=\int\limits_0^\infty {\rm d}r R_{k_f,\,1}(r)R_{0}(r)
\end{equation}
Using the standard quasiclassical wave function and 
assuming $x\ll 1$, we obtain   
\begin{eqnarray}
I_0 &\approx& 
-\frac{1}{4i\bar\eta}\int\limits_0^\infty\frac{d\vartheta}{\cosh^2(\vartheta/2)}
\mbox{e}^{i\xi(\sinh\vartheta+\vartheta)}\nonumber\\
&&\left\{1+i\frac{b_1-b_0}{2\bar\eta}\tanh(\vartheta/2)\right\}\nonumber\\
&=&\frac{\xi}{2\bar\eta}\Bigg(\int\limits_0^\infty d\vartheta \sinh(\vartheta)
\,\mbox{e}^{i\xi(\sinh\vartheta+\vartheta)}\nonumber\\
&&-\frac{i}{\bar\eta}\int\limits_0^\infty d\vartheta 
\,\mbox{e}^{i\xi(\sinh\vartheta+\vartheta)}\Bigg)\,,
\end{eqnarray}
where $b_l=(l+1/2)^2$ and the second expression is obtained after an 
integration by parts.
According to \cite{Dy1999}, the classically forbidden 
part can be incorporated by shifting the integration 
contour for $\vartheta$ into the complex plane, 
via the replacement $\vartheta\to\vartheta+i\pi$. As a result we arrive at 
\begin{eqnarray}\label{M0f}
I_1 &=&-\frac{\xi}{2\bar\eta}\,
\mbox{e}^{-\pi\xi}\Bigg(\int\limits_0^\infty d\vartheta \sinh(\vartheta)
\,\mbox{e}^{i\xi(\sinh\vartheta-\vartheta)}\nonumber\\
&&+\frac{i}{\bar\eta}\int\limits_0^\infty d\vartheta 
\,\mbox{e}^{i\xi(\sinh\vartheta-\vartheta)}\Bigg)\,.
\end{eqnarray}
Similarly, the leading in $1/\bar\eta$ contribution to the quadrupole 
amplitude is determined by the integral [see the second term in 
square brackets in Eq.~(\ref{MN})]
\begin{eqnarray}\label{M21}
I_{21}&=&\int\limits_0^\infty 
{\rm d}r R_{k_f,\,2}(r)r \omega R_{0}(r)\nonumber\\
&=&\frac{iv\xi}{2\bar\eta}\mbox{e}^{-\pi\xi}  \int\limits_0^\infty d\vartheta 
\,\mbox{e}^{i\xi(\sinh\vartheta-\vartheta)}\,.
\end{eqnarray}
The results (\ref{M0f}) and (\ref{M21})
immediately verify Eqs.~(\ref{Our}) and (\ref{OurQ}).

For the second 
contribution to the quadrupole amplitude, see Eq.~(\ref{MN}),
\begin{eqnarray}
I_{22} &=&
\int\limits_0^\infty {\rm d}r R_{k_f,\,2}(r)\frac{9}{\mu r} R_{0}(r),
\label{M22}
\end{eqnarray}
we find that this term
is suppressed by a factor $1/\bar\eta$. 
However, while $I_{21}$ vanishes in the limit of a small photon energy 
$x\to 0$, the contribution $I_{22}$ tends to 
a nonzero constant at $x=0$, i.e., even though $I_{22}$ is
parametrically suppressed by a factor $1/{\bar \eta}$, 
it constitutes the dominant contribution to the 
dipole-quadrupole interference term at very small photon 
energies, due to its distinctive asymptotic behaviour.

A precise calculation of the $I_{22}$ contribution to the 
interference term is unfortunately hampered by the 
fact that the region of integration around $r \sim r_0$
gives an important contribution to the value of $I_{22}$
due to the inverse power of $r$ in the integrand, 
which thus makes the value of $I_{22}$ 
nuclear model-dependent. In the data analysis of our recent 
experiment~\cite{ourexp},
we therefore did not include the parametrically suppressed term 
$I_{22}$. While good overall agreement between experiment and 
theory was obtained in this way (see Fig.~5 of Ref.~\cite{ourexp}), 
the agreement is better at
photon energies $E_\gamma \geq 250\,{\rm keV}$ as compared to
photon energies below this region. As the region of small $E_\gamma$
coincides with the region where the $I_{22}$ term might contribute 
to the dipole-quadrupole
interference term, its significance cannot be completely ruled 
out at present.

\end{document}